\documentclass[debug]{rmaa}

\usepackage{paralist}

\usepackage{psfrag,color}

\usepackage[latin1]{inputenc}
\usepackage{natbib}
\usepackage{longtable}
\usepackage{lscape}
\usepackage{rotating}
\usepackage{multirow}
\usepackage{graphicx}

\title{Gaseous disks in white dwarfs: Properties of the current sample} 

\author{
  L. Saker,\altaffilmark{1} 
  M. G\'omez,\altaffilmark{1,2}
  L. Garc\'\i{}a\altaffilmark{1}}

\altaffiltext{1}{Observatorio Astron\'omico de C\'ordoba, UNC, Argentina.}

\altaffiltext{2}{CONICET, Argentina.}

\shortauthor{Saker, G\'omez, \& Garc\'\i{}a}
\shorttitle{Gaseous disks in white dwarfs}

\fulladdresses{
\item L. Saker, M. G\'omez \& L. Garc\'\i{}a: Observatorio Astronmico de C\'ordoba, Universidad Nacional de C\'ordoba, Laprida 854, 5000 C\'ordoba, Argentina (leilasaker88@unc.edu.ar).

\item M. G\'omez: Consejo Nacional de Investigaciones Cient\'\i{}ficas y T\'ecnicas (CONICET), Argentina.}

\listofauthors{L. Saker, M. G\'omez \& L. Garc\'i{}a}
\indexauthor{Saker, L.}
\indexauthor{G\'omez, M.}
\indexauthor{Garc\'i{}a, L.}

\abstract{In recent years, the number of white dwarfs (WDs) showing infrared (IR) excesses, attributed to circumstellar dust disks, has grown significantly. Gaseous disks have also been detected in some WDs with dust. We obtained optical spectra with Gemini/GMOS for thirteen WDs exhibiting IR excesses to search for gaseous components. No Ca II emission was detected, which suggest the gas phase may be short-lived compared to the dusty stage. By combining our data with literature, we compiled the largest sample of WDs with and without gas disks. We found no significant differences in stellar properties between the two groups, though disks with gas may be, on average, brighter. The detection rate of Ca II emission supports the idea that the gas phase persists for only a fraction of the total disk lifetime. A more definitive assessment will likely require dedicated searches for both gas and dust across large WD samples.}

\resumen{}

\addkeyword{Stars: white dwarfs}
\addkeyword{Stars: circumstellar matter}
\addkeyword{Line: profiles}
\addkeyword{Accretion, accretion disks}

\begin{document}
\maketitle

\section{Introduction}
\label{sec:intro}

Since the first detection of infrared (IR) excesses around a white dwarf \citep[WD,][]{1987Natur.330..138Z,2005ApJ...635L.161R,2009ApJ...693..697R}, the number of WDs with excesses has increased significantly: about 60 WD are now known to have IR excesses. Satellites (or missions) such as Spitzer and WISE, have greatly contributed to increasing this number \citep{2013ApJ...770...21H,2016NewAR..71....9F}. The IR excesses are usually considered observational proofs of the existence of remnant planetary systems around WDs, known as debris disks. In addition, the most widely accepted explanation in the literature for the existence of these dusty disks is the accretion of debris caused by the tidal destruction of smaller rocky bodies that originally formed a planetary system orbiting the current WD \citep{2003ApJ...584L..91J,2012MNRAS.424..333G,2014ApJ...783...79X,2015Natur.526..546V}.

This scenario outlined in the previous paragraph is supported by the fact that all WDs with IR excesses have heavy elements, such as Ca, Mg, Si, Fe, etc. in their atmospheres \citep{2003ApJ...596..477Z,2010ApJ...722..725Z}. The high gravity of cool WDs makes any possible heavy chemical element in the atmosphere sinks on time scales from a few days to a million years \citep{1958whdw.book.....S}. This time is considerably shorter than the cooling time of these objects ($10^8$ - $10^9$ years) and, therefore, these elements cannot be primordial \citep{1986ApJS...61..197P,2006A&A...453.1051K,2009A&A...498..517K}, but rather they have been accreted in a relatively recent time.

In addition to dusty disks, gaseous disks have been found around 21 WDs, all of them bearing debris disks \citep{2006Sci...314.1908G,2007MNRAS.380L..35G,2008MNRAS.391L.103G,2011AIPC.1331..211G,2012MNRAS.421.1635F,2012ApJ...751L...4M,2014MNRAS.445.1878W,2020ApJ...905...56M,2020ApJ...905....5D,2021MNRAS.504.2707G}. These gas disks are recognized by the detection of double-peaked emission lines in the 8600 \AA{}
CaII triplet. Such profiles unambiguously identify the presence of a Keplerian rotating gaseous disk around these stars \citep{1986MNRAS.218..761H}.

 In this contribution, we present our search for a gas disk in a sample of 13 WDs with a dusty disk. For this purpose, in Section 2, we analyze the spectra acquired with GMOS/Gemini.
 In Section 3 we combine our observed sample with those from the literature to derive the largest known subsample of WDs with gas disk to compare stellar and disk properties in both WDs with and without gaseous component. Finally, in Section 4, we present our conclusions.
 
\section{Observations}

\subsection{Target selection and data reduction}
\label{sec:targets}

We have selected a subset of 13 WDs with confirmed IR excesses; the chosen stars that have temperatures between 11880-21450 K (close to the temperature range in which disks of gas have been previously detected\footnote{We caution, however, that this search as well as others \citep[see, for example,][]{2020MNRAS.493.2127M} do not include higher temperature WDs. This is because the sensitivity of detecting metallic lines in the optical declines at high temperatures \citep{2024Ap&SS.369...43B}.}).
We obtained optical spectra with the 8m Gemini telescopes using the \- Gemini Multi-Object Spectrograph \citep[GMOS;][]{2004PASP..116..425H} in longslit mode. The instrument consists of three 6266$\times$4176 Hamamatsu CCDs, separated by two wide gaps, with a field of view (FOV) of $\sim$ 5.5' square.

The CCDs were binned in the spatial direction by a factor of 4 and we used the R831 grating, with a central wavelength setting of 8600 \AA{}. This setup provides a wavelength coverage of 7500-9750 \AA{} at a resolution of 2 \AA{}. To block 2nd-order contamination of the spectra, we used the OG515 filter. To eliminate the gaps located between the CCDs and to improve the S/N, we applied a dithering of $\pm$50 angstroms around the central wavelength. Table \ref{tab:tab1} provides a log of our observations.

The spectra were reduced using standard spectroscopic techniques (i.e. bias-subtraction, flat-fielding, wavelength-calibration, sky-subtraction, and finally the extraction of the 1D spectra) using specific Gemini/GMOS IRAF routines. To perform this procedure, appropriate bias and flat-field frames were obtained from the Gemini Observatory Archive (GOA\footnote{https://archive.gemini.edu/searchform}). For wavelength calibration, we obtained CuAr Arcs lamps at the targets' elevations.

The signal-to-noise ratio (SNR) and detection limits were empirically estimated from the reduced GMOS spectrum. In a line-free region of the continuum, the pixel-to-pixel noise was measured directly from the data. The noise per resolution element was then computed by multiplying the pixel noise by the square root of the full width at half maximum (FWHM) of a representative spectral line, expressed in pixels. This yielded a value of approximately 26.4 ADU due to noise per resolution element.

Assuming a conservative 3$\sigma$ detection threshold, the minimum detectable signal was converted into a minimum detectable equivalent width by normalizing to the continuum level and scaling by both the spectral dispersion (in \AA{}/pixel) and the square root of the FWHM. This provides an empirical estimate of the smallest equivalent width that can be confidently detected in the spectrum. The resulting value is on the order of $\sim$ 0.01 \AA{}, which defines the lower detection limit for weak absorption or emission features in the observed spectral range.

The residuals from telluric and night sky lines were carefully inspected, particularly around the Ca II triplet, and eliminated. The level of contamination from these contributions, in general, turned out to be negligible compared to the given value of the noise per resolution element.

\begin{table*}[!t]\centering
\small
  \begin{changemargin}{-2cm}{-2cm}
  \caption{Log of the Gemini/GMOS observations searching for Ca II triplet emission at WDs with a detected debris dust disk.} \label{tab:tab1}
      \setlength{\tabnotewidth}{0.95\linewidth}
    \setlength{\tabcolsep}{1.2\tabcolsep} \tablecols{6}
 \begin{tabular}{lccccc}
    \toprule
	WD & Program & Date [UT] & Exposure [sec] & Slit [arcsec] & $\lambda_c$ [\AA{}]	\\
    \hline
WD 1448$+$411 & GN-2016A-Q-76  &  2016-06-24      & 900 & 1.00 & 8600-8650 \\
WD 1457$-$086 & GN-2016A-Q-76  &  2016-06-25$/$28 & 900 & 1.00 & 8600-8650 \\
WD 0106$-$328 & GS-2016B-Q-21  &  2016-10-10      & 360 & 0.75 & 8550-8600-8650  \\
WD 0420$-$731 & GS-2016B-Q-21  &  2016-10-09$/$10 & 360 & 0.75 & 8550-8600-8650  \\
GD 40         & GS-2016B-Q-21  &  2016-10-09      & 360 & 0.75 & 8550-8600-8650  \\
GD 61         & GN-2016B-Q-56  &  2016-11-03      & 250 & 0.75 & 8550-8600-8650  \\
WD 1046$-$017 & GN-2017A-Q-38  &  2017-04-15$/$16 & 600 & 0.75 & 8550-8600-8650  \\
WD 1150$-$153 & GN-2017A-Q-38  &  2017-04-16      & 600 & 0.75 & 8550-8600-8650  \\
WD 1541$+$650 & GN-2017A-Q-38  &  2017-04-16      & 450 & 0.75 & 8550-8600-8650  \\
GD 16         & GS-2017B-Q-74  &  2017-08-27      & 400 & 1.00 & 8550-8600-8650 \\
              &                &  2017-10-18      &        &      &                 \\
GD 56         & GS-2017B-Q-74  &  2017-08-26      & 450 & 1.00 & 8550-8600-8650   \\
WD 0836$+$404 & GN-2018A-Q-307 &  2018-01-20      & 645 & 1.00 & 8550-8600-8650   \\
WD 1116$+$026 & GN-2018A-Q-307 &  2018-01-20      & 600 & 1.00 & 8550-8600-8650   \\
		\hline
	\end{tabular}
	  \end{changemargin}
\end{table*}

\subsection{Search of emission lines of Ca II triplet}
\label{sec:ca} 

\begin{figure}[!t]\centering
\includegraphics[width=\columnwidth]{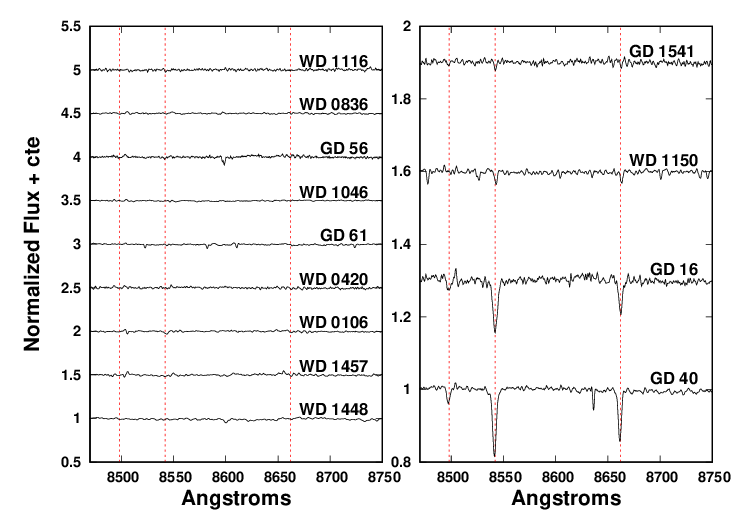}
    \caption{Gemini/GMOS spectra around de 8600 \AA{} Ca II triplet for our observed sample (see Table \ref{tab:tab1}).  The normalized fluxes in the vertical scale are arbitrary shifted to better visualize the complete sample. The dashed red lines indicate the position of the Ca II triplet.}
    \label{fig:espectros}
\end{figure}

We visually inspected the region between 8470-8750 \AA{} of our spectra in order to search for the emission lines of the Ca II triplet. Figure \ref{fig:espectros} shows this region for each spectrum. For the WDs in the left panel, the Ca II triplet appears in neither emission nor absorption. In the right panel, we detect the Ca II triplet in absorption for GD 40 and GD 16. \citet{2010ApJ...709..950K} and \citet{2017ApJ...849...77D} also detected the triplet of Ca II in absorption in GD 40. WD 1150$-$153 and WD 1541$+$650 also have these lines in absorption, but with less intensity. 

For WD 0420$-$731, WD 1046$-$017, WD 0836$+$404 and WD 1448$+$411 there are no previously published spectra covering the Ca II triplet region. \citet{2017ApJ...849...77D} did not find the Ca II triplet lines for WD 0110$-$565 and WD 0106$-$328. \citet{2020MNRAS.493.2127M} obtained spectra of 20 WDs with an IR excess, 9 of which are included in our sample. No WD in their sample exhibits the calcium triplet in emission. 

\section{Characteristics of white dwarfs with gaseous disks}

In this section, we compare and analyze properties of different subsamples of WDs. To obtain the stellar parameters (Table \ref{tab:full1}), we used the Montreal White Dwarf Database\footnote{http://montrealwhitedwarfdatabase.org/.} \citep[MWDD,][]{2017ASPC..509....3D}. To ensure that the analyzed samples are as homogeneous as possible, we used the work of  \citet{2024A&A...682A...5V} to obtain the T$_{eff}$ and log g values.  These parameters were applied to determine the total ages of the stars using the {\it Wdwarfdate} open-source Python package developed by \citet{2022AJ....164...62K}. {\it Wdwarfdate} provides main-sequence (MS), cooling and total (MS $+$ cooling) ages, as well as initial and final masses. The cooling ages and final masses derived from the code agree (within the errors) with those reported in the literature, indicating the consistency of our samples.

The 2MASS H (1.65 $\mu$m), K (2.17 $\mu$m) band data, WISE W1 (3.35 $\mu$m), W2 (4.60 $\mu$m) magnitudes and Gaia Data Release 2 (DR2) G$_{BP}$ (0.513 $\mu$m), G$_{RP}$ (0.778 $\mu$m) passbands\footnote{Central wavelengths of Gaia passbands are from \citet{2018A&A...617A.138W}.} were employed to derive color excesses in the IR, attributed to the dust disks (Table \ref{tab:full2}). The disk fractional luminosities (${\rm {L_{IR}/L_*}}$) were obtained from \citet{2015MNRAS.449..574R} and \citet{2020ApJ...905....5D}. The samples are comparable as they were derived using the same methodology \citep{2020ApJ...905....5D}. In all cases, WDs with IR excess at near-IR (Spitzer) wavelengths are selected following the usual criteria, using flux excesses ($\chi$) defined as:
\begin{equation}
    \chi = \frac{F_{\mathrm{obs}} - F_{\mathrm{model}}}{\sqrt{\sigma_{\mathrm{obs}}^2 + \sigma_{\mathrm{model}}^2}},
\end{equation}
where $F_{\mathrm{obs}}$ and $F_{\mathrm{model}}$ are the observed and model-predicted photospheric fluxes, respectively, and $\sigma_{\mathrm{obs}}$ and $\sigma_{\mathrm{model}}$ are their associated uncertainties. Significant excesses are defined as those with $\chi > 3$, which corresponds to a $3\sigma$ confidence level.

 \subsection{Comparison of WDs with gaseous disks with other subsamples of WDs }

In order to analyze the existence of gaseous disks in WDs with dusty disks over the largest sample available up-to date,  we combined our sample (see Section \ref{sec:ca}) with those previously reported in the literature \citep[see for example][]{2006Sci...314.1908G,2007MNRAS.380L..35G,2020MNRAS.493.2127M,2020ApJ...905...56M,2023ApJ...944...23W}. The final complete sample is presented in the Appendix (Table \ref{tab:full1}). We divided this final sample into two subsamples: those with and those without gas disks, depending on the detection of the Ca II triplet in emission in their spectra. It is worth mentioning that all stars in Table \ref{tab:full1} were searched for the Ca II triplet.

We included two other subsamples in our analysis. WDs with contaminated atmospheres\footnote{Contaminated or Polluted WDs, in which planetary debris has contaminated otherwise pristine H or He atmospheres, 
are identified based on the detection of heavy metallic (magnesium, iron, calcium, etc.) lines in their spectra \citep{2014A&A...566A..34K}.} but without IR excess, and naked WDs (without/unknown contamination and IR excess). Data for these latter two groups were obtained from only two sources: \citet{2009ApJ...694..805F} and \citet{2019MNRAS.487..133W}. It is worth noting that the WDs present in both independent studies were classified into the same category --either with or without/unknown contamination or IR excess-- in both works. 

\begin{figure*}
\centering
\makebox[\textwidth][c]{%
\begin{tabular}{c}
    \includegraphics{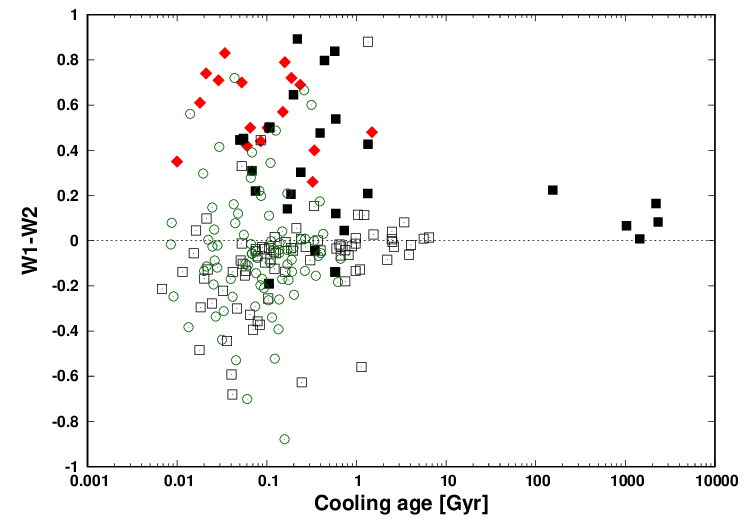} \\
    \includegraphics{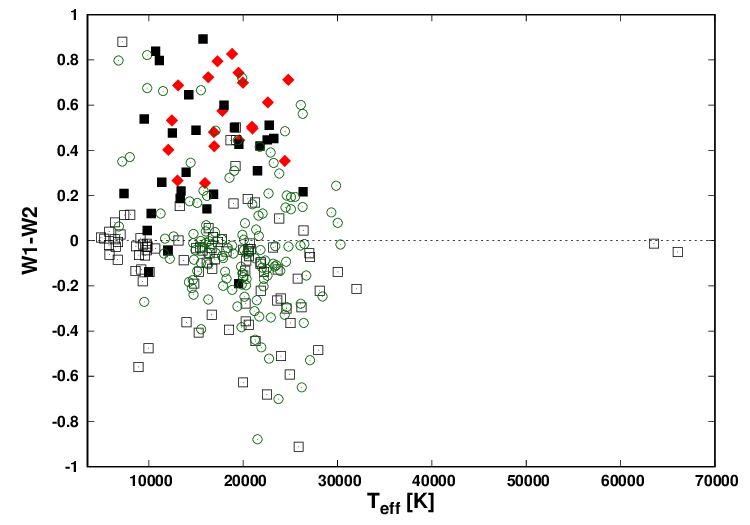}
    \end{tabular}
    }
    \caption{W1$-$W2 vs. age (upper panel) and T$_{eff}$ (lower panel). The filled red diamonds correspond to WD with debris and gas disk, the filled black squares to WD with only debris disk, the empty black squares to WD with contaminated atmospheres without IR excess, and empty green circles represent naked WD (without/unknown contamination and IR excess).}
    \label{fig:comp}
\end{figure*}
 
Figure \ref{fig:comp} shows W1$-$W2 vs cooling age (upper panel) and T$_{eff}$ (lower panel) for four subsamples: The filled red diamonds correspond to WDs with debris and gas disk, the filled black squares to WDs with only debris disk, the empty black squares to WDs with contaminated atmospheres without IR excess, and empty green circles represent naked WDs. In these plots, WDs with 2MASS and WISE uncertain photometry were not included. We note that the four subsamples span similar ranges in cooling age and/or in T$_{\rm eff}$. The only exception are two hot WDs with contaminated atmospheres, with T$_{\rm eff}$ $>$ 60000 K. WDs with dusty disks with and without the gas component have a large color excess (median W1$-$W2 $=$ 0.53 and 0.48), suggesting that the dust component is similar in disks with or without gas. As expected, WDs without dusty disks (with or without/unknown contaminated atmospheres) do not show a color excess (median W1$-$W2 $=$ $-$0.137 and $-$0.168, respectively). 

\subsection{Are WDs with gaseous disks different from those with disks but without a gas component?}\label{section3.2}

\begin{figure*}
\centering
\makebox[\textwidth][c]{%
    \begin{tabular}{c c}
    \includegraphics[width=7.5cm,height=6cm]{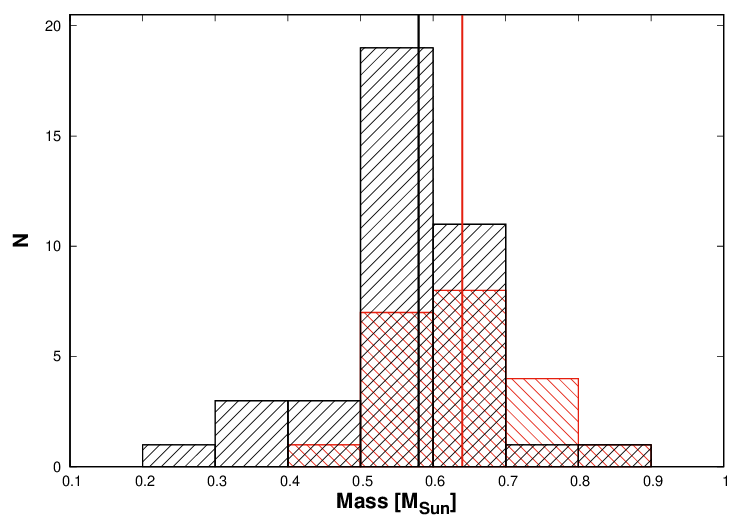}
    \includegraphics[width=7.5cm,height=6cm]{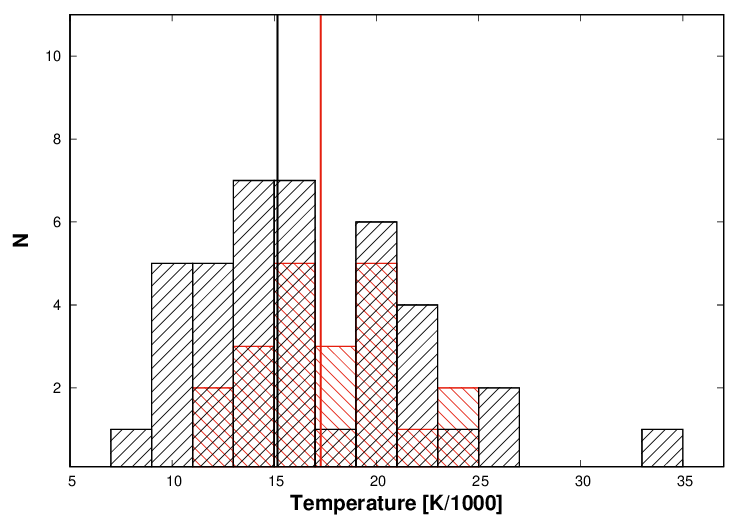} \\
    \includegraphics[width=7.5cm,height=6cm]{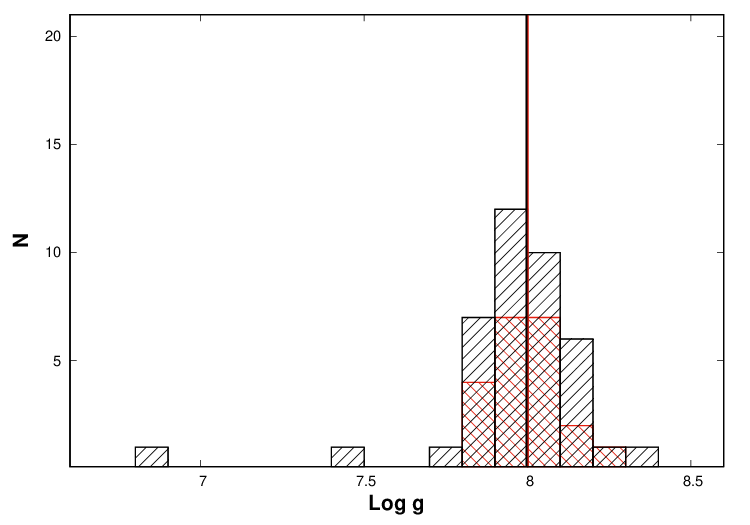}
    \includegraphics[width=7.5cm,height=6cm]{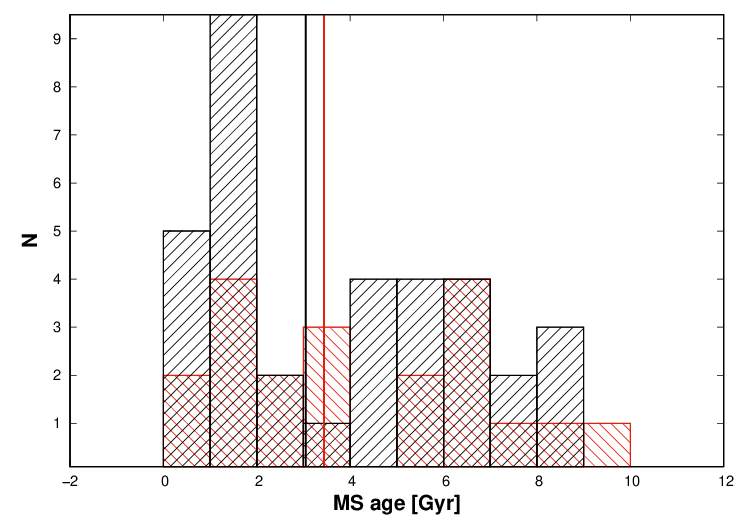}\\
    \includegraphics[width=7.5cm,height=6cm]{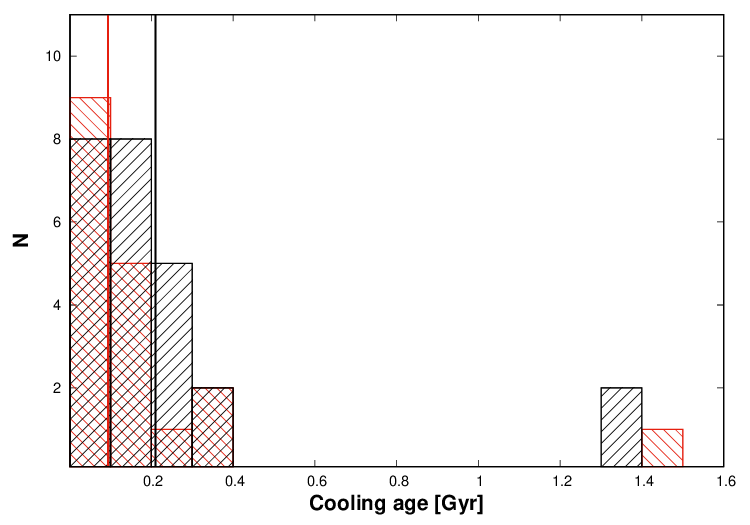}
    \includegraphics[width=7.5cm,height=6cm]{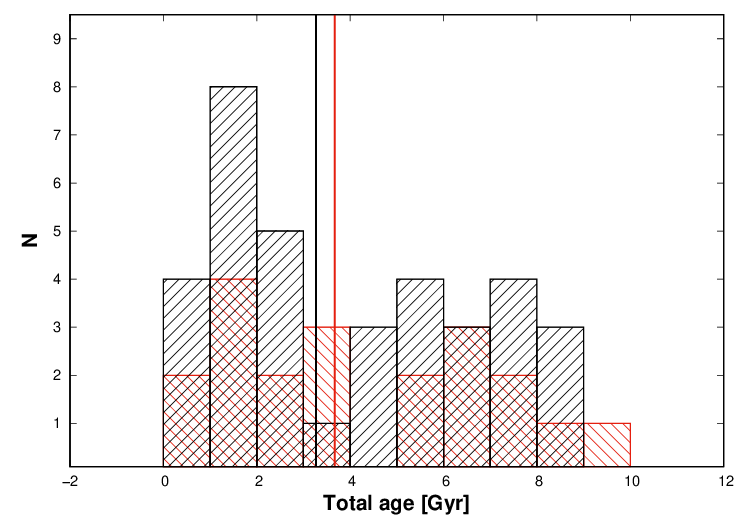}
\end{tabular}
}
\caption{Histograms of the stellar parameters mass, temperature, log g, and cooling time obtained from the literature, as well as main sequence age and total age derived using {\it wdwarfdate}, for the analyzed subsamples: stars with (red) and without (black) disk gas detected. The red and black lines correspond to the median of each distribution.}
    \label{fig:hist1}
\end{figure*}

\begin{figure*}
\centering
\makebox[\textwidth][c]{%
\begin{tabular}{c c}
     \includegraphics[width=7.5cm,height=6cm]{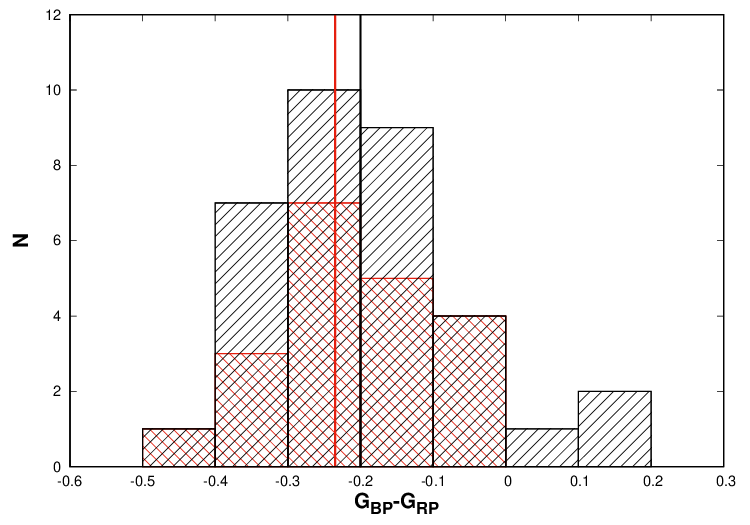}
    \includegraphics[width=7.5cm,height=6cm]{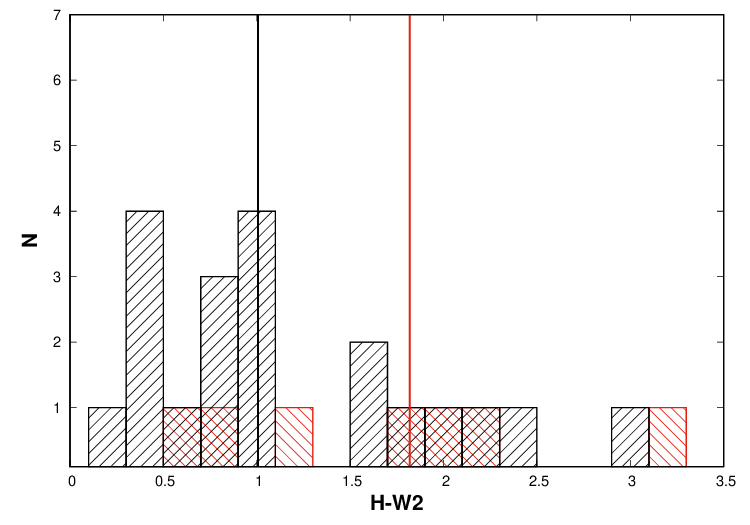}\\
    \includegraphics[width=7.5cm,height=6cm]{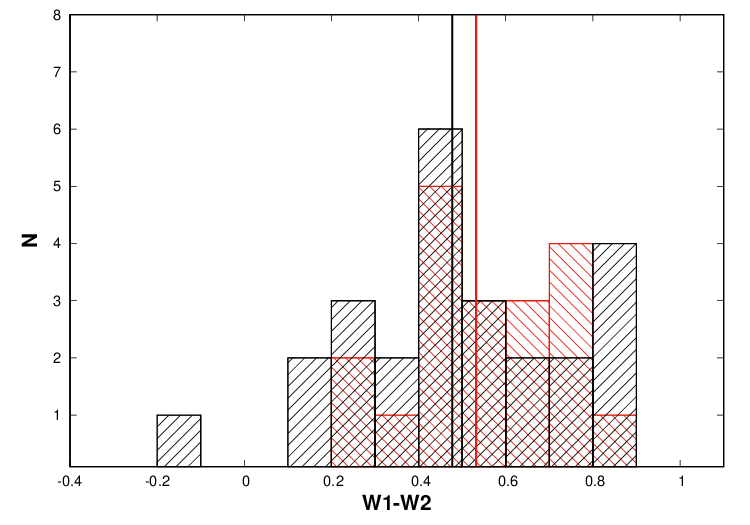}
     \includegraphics[width=7.5cm,height=6cm]{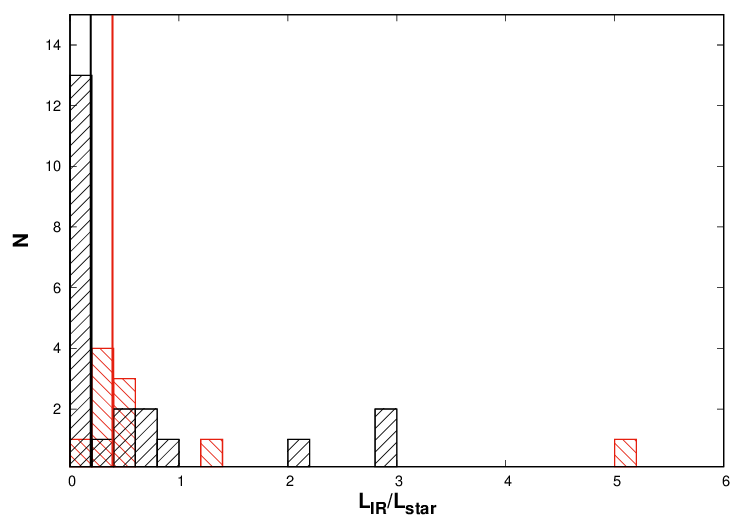}
    \end{tabular}
    }
    \caption{Histograms of color excess G$_{BP}-$G$_{RP}$, H$-$W2, and W1$-$W2, as well as the fractional infrared luminosity ${\rm {L_{IR}/L_*}}$, for the analyzed subsamples: stars with (red) and without (black) disk gas detected. The red and black lines correspond to the median of each distribution.}
    \label{fig:hist2}
\end{figure*}

We compared stellar properties of WDs with IR excesses with and without gaseous components, searching for hints that may indicate why some WDs have gas disks while others lack this component. On the other hand, we inspect the colors G$_{BP}-$G$_{RP}$, H$-$W2, W1$-$W2, and the fractional infrared luminosity (${\rm {L_{IR}/L_*}}$), as they may suggest the presence of warmer or brighter dust components in the disks. This could hint to different properties in the disks with and without gas content. Figure \ref{fig:hist1} shows the distribution of stellar parameters, whereas in Figure \ref{fig:hist2}, the distributions of colors and fractional infrared luminosities of dusty disks, with and without the gas component, are displayed. The medians of all the parameters analyzed are shown in Table \ref{tab:medians}.

\begin{table}[!t]\centering
\small
\setlength{\tabnotewidth}{0.5\columnwidth}
\tablecols{3}
\setlength{\tabcolsep}{2.8\tabcolsep} 
    \caption{Medians of the stellar parameters and \\color excesses related to the disk.}
    \begin{tabular}{l c c}
    \toprule
     Parameters & \multicolumn{2}{c}{Medians} \\ \cmidrule{2-3}
                 & With gas disk & Without gas disk \\
    \midrule
    Mass [M$_\odot$]  & 0.640   & 0.580 \\
    T$_{eff}$ [K]  &  17284   & 15156\\
    Log g             &  8    & 7.996 \\
    MS age [Gyr]   & 3.440 & 3.050  \\
    Cooling age [Gyr]         & 0.094   & 0.210 \\
    Total age [Gyr] & 3.670 & 3.270 \\
     G$_{BP}-$G$_{RP}$ & $-$0.235  & $-$0.200 \\
    H$-$W2             & 1.819    & 1.006 \\
    W1$-$W2           & 0.532   & 0.477 \\
    $L_{IR}/L_*$ & 0.395   & 0.190        \\
    \bottomrule   
    \end{tabular}
    \label{tab:medians}
\end{table}

Since these histograms do not have a normal distribution (i.e., a clear bell shape), we used the Mann-Whitney U test (a non-parametric statistical method) to determine the null hypothesis (H$_0$) that our subsamples are drawn from the same parent population. This statistical test compares the ranks of randomly selected values from sample 1 and sample 2, and determines the probability that one sample tends to have higher values than the other.
Table \ref{tab:test} summarizes the values obtained when applying the test. In general, the p-value represents the probability of rejecting a correct H$_0$, and a significance level of $\alpha=$0.05 is assumed. When p$< \alpha$, H$_0$ can be rejected with a low chance of error. Formally, the p-values of stellar parameters (mass, temperature, cooling ages and Log g) are smaller than 0.05. However, we can not claim that the samples are statistically different given  the proximity of these p-values with a threshold of 0.05 chosen.

On the other hand, even though the distributions of colors and fractional infrared luminosity, ${\rm {L_{IR}/L_*}}$ are statistically similar or indistinguishable (see Table \ref{tab:test}), the medians of infrared colors excesses
(particularly in H$-$W2) and ${\rm {L_{IR}/L_*}}$ are significantly larger for WDs with a gas component than for WDs without gas (see Table \ref{tab:medians}). This suggests that dusty disks are brighter (or more massive) in WDs with gas than in those without gas component detected. However, we caution that this trend is based on small samples and therefore it needs to be confirmed as more observations become available.

\begin{table}[!t]\centering
\small
\setlength{\tabnotewidth}{0.5\columnwidth}
\tablecols{3}
\setlength{\tabcolsep}{2.8\tabcolsep} 
    \caption{The Mann-Whitney U-Test applied \\to the analyzed subsamples.}
    \begin{tabular}{l c}
\toprule
  Parameters &   $p-value$   \\
 \midrule
Mass         & 0.04036 \\  
Temperature  & 0.07215 \\   
Log g        & 0.35942 \\ 
 MS age & 0.23576 \\
Cooling age  & 0.0164 \\  
Total age & 0.27093\\
G$_{BP}-$G$_{RP}$ & 0.5905 \\
H$-$W2     & 0.1795 \\ 
W1$-$W2     & 0.3937 \\ 
$L_{IR}/L_*$ &  0.11184 \\
\bottomrule
    \end{tabular}
    \label{tab:test}
\end{table}

\section{Summary and conclusions}

We obtained Gemini/GMOS optical spectra of thirteen WDs with IR excesses to search for the gaseous counterparts to the dust disks based on the Ca II double-peak emission profile. No such profile was detected in any WD in our sample. 

Although the origin of the gas component in WDs' disks is not reliably known, three mechanisms have been proposed: 1) sublimation of the dust particles at the inner edge of the debris disk \citep{2012MNRAS.423..505M}, 2) vaporization of small solid particles originated in a collisional cascade \citep{2017ApJ...844..116K,2017ApJ...850...50K}, 3) the existence of a planetesimal in a close-in orbit within the debris disk. The material in this body is sublimated by the radiation of the central WD generating gas component \citep{2019Sci...364...66M}. In particular, these authors propose the presence of a highly eccentric planetesimal, whose passages through the orbital periastrum would explain the variable (but periodic, P $\sim$2h) emission found in the CaII triplet of the WD SDSS J1228$+$1040.

While the frequency of WDs with heavy elements in their atmospheres is 25-50\% \citep{2003ApJ...596..477Z,2010ApJ...722..725Z,2014A&A...566A..34K}, several studies have established that only 1-4\% of the WDs show IR excesses \citep{2007ApJS..171..206M,2009ApJ...694..805F,2012ApJ...760...26B,2015MNRAS.449..574R,2019MNRAS.489.3990R,2019MNRAS.487..133W}. \citet{2020MNRAS.493.2127M} provide the first statistical study of the frequency of gaseous components in WDs with debris disks. They estimated that only $\sim$4\% of the dusty disk WDs also show evidence of a gaseous component. Our inability to detect the CaII triplet in the observed sample of thirteen WDs is consistent with a low frequency of gaseous disks.

We combined our observed WDs with other objects from the literature to analyze the largest available sample of WDs with and without a gaseous disk. Our results indicate that there are no significant differences between WDs with and without gaseous disks in stellar properties. Regarding the properties of the disks, Table  \ref{tab:full1} lists the 61 WDs with confirmed dust disks for which searches for gas disks have also been carried out. Evidence of a gas disk is found in 21 of these systems ($\sim$1/3 of the sample). This could suggest that gas disks have shorter lifetimes compared to dust disks. Furthermore, the median of the ${\rm {L_{IR}/L_*}}$ ratio is larger for  WDs with gas disk than for WDs without the gas component (Figure \ref{fig:hist2}, lower right panel). This may imply that gas is predominantly present during the early stages of disk evolution and that its production decreases as the disk evolves and becomes more dynamically stable \citep{2017ApJ...850...50K}.

However,  there are important caveats to consider. If the presence of a dust disk is a necessary condition for the formation of a gas disk, then estimating the gas disk fraction from a sample pre-selected to have dust disks may introduce a bias that favors detecting gas disks. In addition, both subsamples in Table \ref{tab:full1} are relatively small. Taking these factors into account, the fraction of $\sim$1/3 should be considered with caution and, at most, as an upper limit to the true occurrence rate of gas disks among dusty disk WDs. In any event, this is a different approach from that proposed by \citet{2020MNRAS.493.2127M}, who estimated the frequency of gas disks using samples of WDs selected based on magnitude and/or minimum S/N requirements, regardless of the previous existence of known dust disks. From these samples, these authors derived a frequency of $\sim$4\%, as mentioned above. A more definitive assessment will most likely require dedicated searches for gas and dust across large WD samples.

\section*{Acknowledgments}

We thank the referee for their thoughtful and thorough review of our manuscript. We are particularly grateful for the detailed comments and suggestions that improved the content and clarity of the work.

L.S. gratefully acknowledges the support from doctoral and postdoctoral fellowships from CONICET (Argentina), which made it possible to carry out the majority of the work presented in this paper.

Based on observations obtained at the international Gemini Observatory, a program of NSF NOIRLab, which is managed by the Association of Universities for Research in Astronomy (AURA) under a cooperative agreement with the U.S. National Science Foundation on behalf of the Gemini Observatory partnership: the U.S. National Science Foundation (United States), National Research Council (Canada), Agencia Nacional de Investigaci\'{o}n y Desarrollo (Chile), Ministerio de Ciencia, Tecnolog\'{i}a e Innovaci\'{o}n (Argentina), Minist\'{e}rio da Ci\^{e}ncia, Tecnologia, Inova\c{c}\~{o}es e Comunica\c{c}\~{o}es (Brazil), and Korea Astronomy and Space Science Institute (Republic of Korea).

\begin{appendices}
\section{Appendix I} %
\label{sec:ap-A}

\begin{longtable}{ccccccc}
\caption{Full sample of white dwarfs with IR excesses attributed to debris disks, ordered by T$_{eff}$. Stellar parameters from MWDD.} \label{tab:full1} \\

\hline 
\multicolumn{1}{c}{WD} & 
\multicolumn{1}{c}{T$_{eff}$} &
\multicolumn{1}{c}{Log g} &
\multicolumn{1}{c}{Mass} &
\multicolumn{1}{c}{Cooling age} &
\multicolumn{1}{c}{Gas disk?} &
\multicolumn{1}{c}{Ref.} \\ 

\multicolumn{1}{c}{} & 
\multicolumn{1}{c}{[K]} &
\multicolumn{1}{c}{} &
\multicolumn{1}{c}{[M$_\odot$]} &
\multicolumn{1}{c}{[Gyr]} &
\multicolumn{1}{c}{} &
\multicolumn{1}{c}{} \\ 

\hline 
\endfirsthead

\multicolumn{7}{c}%
{{\tablename\ \thetable{} Stellar parameters of the full sample of white dwarfs with debris disks (cont.)}} \\
\hline 
\multicolumn{1}{c}{WD} & 
\multicolumn{1}{c}{T$_{eff}$} &
\multicolumn{1}{c}{Log g} &
\multicolumn{1}{c}{Mass} &
\multicolumn{1}{c}{Cooling age} &
\multicolumn{1}{c}{Gas disk?} &
\multicolumn{1}{c}{Ref.} \\

\multicolumn{1}{c}{} & 
\multicolumn{1}{c}{[K]} &
\multicolumn{1}{c}{} &
\multicolumn{1}{c}{[M$_\odot$]} &
\multicolumn{1}{c}{[Gyr]} &
\multicolumn{1}{c}{} &
\multicolumn{1}{c}{} \\ 

\hline 
\endhead

\hline
\endfoot

\hline \hline
\endlastfoot

WD 1455$+$298       &  7366   &  7.98  &  0.59  &  1.34  &    no   &   4, 30 \\
WD 2115$-$560       &  9518   &  7.96  &  0.58  &  0.59  &    no   &   6, 36 \\
HE 2221$-$1630      &  9824   &  8.11  &  0.67  &  0.73  &    no   &   4, 23 \\
HS 0307$+$0746      &  9985   &  7.98  &  0.59  &  0.58  &    no   &   4, 23 \\
WD 1225$-$079       &  10261  &  7.89  &  0.58  &  0.59  &    no   &   22, 29 \\
WD 1729$+$371       &  10717  &  8.2   &  0.55  &  0.57  &    no   &   4, 33 \\
WD 2326$+$049       &  11117  &  8     &  0.62  &  0.44  &    no   &   4, 37 \\
WD 0836$+$404       &  11365  &  7.99  &  0.71  &   -    &    no   &   1, 2 \\
WD J1221$+$1245     &  12012  &  8.09  &  0.57  &  0.35  &    no   &   13 \\
SDSS J1617$+$1620   &  12073  &  7.98  &  0.61  &  0.34  &    yes  &   9, 11, 18 \\
WD J0959$-$0200     &  12456  &  7.91  &  0.64  &   -    &    yes  &   12, 13 \\
WD 1116$+$026       &  12503  &  8.06  &  0.64  &  0.39  &    no   &   1, 4, 7 \\
SDSS J0842$+$2948   &  12771  &  7.71  &  0.44  &  0.24  &    no   &  42 \\
WD 0145$+$234       &  13049  &  8.09  &  0.67  &   -    &    yes  &   38, 39, 40 \\
WD J1930$-$5028     &  13104  &  8.02  &  0.55  &  0.24  &    yes  &   38, 39 \\
WD 2132$+$096       &  13323  &  8     &  0.59  &   -    &    no   &   4, 35 \\
WD 1554$+$094       &  13428  &  6.87  &  0.22  &  0.07  &    no   &   13, 24, 32 \\
WD 1046$-$017       &  13821  &  8     &  0.58  &  0.24  &    no   &   1, 2 \\
WD 1448$+$411       &  13959  &  7.95  &  0.57  &  0.24  &    no   &   1, 2 \\
SDSS J1238$+$2910   &  14183  &  8.35  &  0.86  &  0.43  &    no   & 42  \\
WD 2207$+$121       &  14235  &  7.88  &  0.57  &  0.2   &    no   &   24, 26 \\
SDSS J0738$+$1835   &  14765  &  8.16  &  0.84  &   -    &    yes  &   8, 9 \\
Gaia 2204$+$0233    &  14817  &  8.08  &  0.66  &  0.17  &    no   & 42   \\
WD 1145$+$017       &  15007  &  8.06  &   -    &   -    &    no   &   27, 28 \\
WD 0902$+$399       &  15089  &  7.46  &  0.33  &  0.09  &    no   &  42 \\
Gaia 0155$+$0431    &  15223  &  7.99  &  0.58  &  0.22  &    no   &  42 \\
WD 1551$+$175       &  15290  &  7.98     &  0.57  &  0.17  &    no   &   26, 31 \\
GD 56               &  15744  &  8.07  &  0.59  &  0.22  &    no   &   1, 4, 7 \\
SDSS J0234$-$0406   &  15949  &  8.25  &  0.71  &  0.32  &    yes  &   38, 39 \\
HE 0106$-$3253      &  16156  &  8.04  &  0.62  &  0.17  &    no   &   1, 4, 5 \\
WD 0842$+$572       &  16293  &  8     &  0.51  &  0.19  &    yes  &   38, 39, 40 \\
GD 61               &  16868  &  8.12  &  0.62  &  0.19  &    no   &   1, 4, 6 \\
HE 1349$-$2305      &  16891  &  7.97  &  0.67  &  1.49  &    yes  &   5, 17 \\
SDSS J0347$+$1624   &  16927  &  7.82  &  0.69  &  0.06  &    yes  &   38, 39, 40, 41 \\
SDSS J1043$+$0855   &  16929  &  8.03  &  0.57  &  0.14  &    yes  &   11, 14 \\
Gaia J0611$-$6931   &  17284  &  7.95  &  0.74  &  0.16  &    yes  &   38, 39, 40, 41 \\
Gaia J0644$-$0352   &  17800  &  8.04  &  0.72  &  0.15  &    yes  &   38, 39, 40, 41 \\
WD 0420$-$731       &  17963  &  7.95  &  0.58  &   -    &    no   &   1, 2 \\
WD J0529$-$3401     &  18821  &  7.82  &  0.64  &  0.03  &    yes  &   38, 39 \\
WD 1015$+$161       &  19074  &  8.02  &  0.64  &  0.11  &    no   &   4, 7, 19 \\
SDSS J1058$+$0206   &  19268  &  7.95  &  0.49  &  0.10  &    no   & 42 \\
WD J2212$-$1352     &  19505  &  7.93  &  0.58  &  0.02  &    yes  &   38, 39 \\
WD 1457$-$086       &  19515  &  7.88  &  0.38  &  0.11  &    no   &   1, 3, 4 \\
WD 0110$-$565       &  19546  &  8.11  &  0.67  &  1.35  &    no   &   4, 5, 6 \\
SDSS J0845$+$2257   &  19583  &  8.06  &  0.58  &  0.09  &    yes  &   10, 11 \\
PG 1031$+$063       &  19736  &  7.86  &  0.51  &  0.07  &    no   & 42 \\
SDSS J1228$+$1040   &  19970  &  8.04  &  0.73  &  0.05  &    yes  &   15, 16 \\
WD 2328$+$107       &  20011  &  7.85  &  0.38  &  0.08  &    no   & 42 \\ 
Gaia J0510$+$2315   &  20938  &  8.17  &  0.65  &  0.1   &    yes  &   38, 39, 40 \\
WD 1622$+$587       &  20986  &  7.95  &  0.52  &  0.07  &    yes  &   38, 39, 40, 41 \\
WD 1018$+$410       &  21520  &  8.07  &  0.57  &  0.07  &    no   &   13, 26 \\
WD 0420$+$520       &  21780  &  8.05  &  0.69  &   -    &    no   &   4, 20 \\
WD 0843$+$516       &  22555  &  7.85  &  0.47  &  0.05  &    no   &   4, 25 \\
SDSS J0006$+$2858   &  22621  &  7.9   &  0.64  &  0.02  &    yes  &   38, 39, 40 \\
EC 05365$-$4749     &  22758  &  8.13  &  0.55  &   -    &    no   &   4, 24 \\
WD 1929$+$011       &  23240  &  8.09  &  0.6   &  0.06  &    no   &   13, 34, 35 \\
WD J2133$+$2428     &  24408  &  7.84  &  0.57  &  0.01  &    yes  &   38, 39 \\
Gaia J2100$+$2122   &  24790  &  8.04  &  0.49  &  0.03  &    yes  &   38, 39, 40, 41 \\
PG 0010$+$281       &  26354  &  7.85  &  0.57  &   -    &    no   &   20, 21, 22 \\
SDSS J0928$+$1332   &  26401  &  8.21  &  0.67  &   -    &    no   & 42 \\
WD 1454$+$172       & 34518   &  8.19  &  0.63  &  0.01       & no   & 42 \\
\end{longtable}
\vspace{-5mm}
\begin{changemargin}{0cm}{-2cm}
\noindent References: 1) This work, 2) \citet{2013ApJ...770...21H}, 3) \citet{2009ApJ...694..805F}, 4) \citet{2020MNRAS.493.2127M}, 5) \citet{2017ApJ...849...77D}, 6) \citet{2012ApJ...749..154G}, 7) \citet{2007ApJ...663.1285J}, 8) \citet{2010ApJ...719..803D}, 9) \citet{2011AIPC.1331..211G}, 10) \citet{2008MNRAS.391L.103G}, 11) \citet{2012ApJ...750...86B}, 12) \citet{2011MNRAS.417.1210G}, 13) \citet{2012MNRAS.421.1635F}, 14) \citet{2007MNRAS.380L..35G}, 15) \citet{2006Sci...314.1908G}, 16) \citet{2009ApJ...696.1402B}, 17) \citet{2012ApJ...751L...4M}, 18) \citet{2014MNRAS.445.1878W}, 19) \citet{2012MNRAS.424..333G}, 20) \citet{2016MNRAS.459.1415B}, 21) \citet{2015ApJ...806L...5X}, 22) \citet{2019AJ....158..242X}, 23) \citet{2010ApJ...714.1386F}, 24) \citet{2016ApJ...831...31D}, 25) \citet{2012ApJ...745...88X}, 26) \citet{2015MNRAS.448.2260G}, 27) \citet{2015Natur.526..546V}, 28) \citet{2016ApJ...816L..22X}, 29) \citet{2011ApJ...741...64K}, 30) \citet{2008ApJ...674..431F}, 31) \citet{2019ApJ...872L..25D}, 32) \citet{2003ApJ...596..477Z}, 33) \citet{2007AJ....133.1927J}, 34) \citet{2010MNRAS.404L..40V}, 35) \citet{2014MNRAS.444.2147B}, 36) \citet{2019MNRAS.490..202S}, 37) \citet{1987Natur.330..138Z}, 38) \citet{2019MNRAS.482.4570G}, 39)
\citet{2021MNRAS.504.2707G}, 40) \citet{2020ApJ...905...56M}, 41) \citet{2020ApJ...905....5D}, 42) \citet{2023ApJ...944...23W}.
\end{changemargin}

\begin{longtable}{@{\extracolsep{\fill}}ccccc}
\caption{Colors and the fractional infrared luminosity ${\rm {L_{IR}/L_*}}$ of white dwarfs with debris disks (confirmed and candidates).} \label{tab:full2} \\
\hline
WD & G$_{BP}-$G$_{RP}$ & H$-$W2 & W1$-$W2 & ${L_{IR}/L_*}$ \\
\hline
\endfirsthead

\multicolumn{5}{c}%
{{\tablename\ \thetable{} Colors and ${\rm {L_{IR}/L_*}}$ of white dwarfs with debris disks (cont.)}} \\
\hline
WD & G$_{BP}-$G$_{RP}$ & H$-$W2 & W1$-$W2 & ${L_{IR}/L_*}$ \\
\hline
\endhead

\hline
\endfoot

\hline \hline
\endlastfoot

WD 1455$+$298$^{\ast \ast}$ & 0.45(0.01)             & 0.35(0.11)             & 0.21(0.06)             &  0.19 \\
WD 2115$-$560$^{\ast \ast}$ & 0.17(0.01)             & 0.90(0.08)             & 0.54(0.05)             &  0.84 \\
HE 2221$-$1630              & 0.16(0.01)             & 0.38(0.26)             & 0.05(0.17)$^{\ast}$    &  0.67 \\
HS 0307$+$0746              & 0.13(0.01)             & $-$0.23(0.45)$^{\ast}$ & $-$0.14(0.27)$^{\ast}$ &  0.18 \\
WD 1225$-$079               & $-$0.01(0.01)$^{\ast}$ & 0.19(0.14)             & 0.12(0.10)             &  0.05 \\
WD 1729$+$371               & 0.09(0.01)             & -                      & 0.84(0.07)             &  2.02 \\
WD 2326$+$049               & 0.02(0.02)$^{\ast}$    & 2.33(0.05)             & 0.80(0.05)             &  2.87 \\
WD 0836$+$404               & 0.01(0.01)$^{\ast}$    & 0.33(0.23)             & 0.26(0.14)             &  - \\
WD J1221$+$1245             & $-$0.01(0.04)$^{\ast}$ & 1.93(0.37)             & $-$0.04(0.34)$^{\ast}$ &  - \\
SDSS J1617$+$1620           & $-$0.07(0.01)          & -                      & 0.40(0.17)             &  1.28 \\
WD J0959$-$ 0100 & 0.08(0.08)$^{\ast}$    & 1.82(0.34)             & 0.53(0.36)             &  - \\
WD 1116$+$026               & $-$0.03(0.01)          & 0.97(0.09)             & 0.48(0.07)             &  0.48 \\
SDSS J0842$+$2948           & $-$0.05(003)           & -                      & 0.60(0.15)             &  - \\
WD 0145$+$234               & $-$0.07(0.01)          & 0.70(0.09)             & 0.27(0.06)             &  - \\
WD J1930$-$5028             & $-$0.08(0.02)          & -                      & 0.69(0.15)             &  - \\
WD 2132$+$096               & $-$0.08(0.01)          & $-$0.11(0.38)$^{\ast}$ & 0.19(0.25)$^{\ast}$    &  0.11 \\
WD 1554$+$094               & $-$0.29(0.05)          & 2.26(0.46)             & 0.22(0.43)$^{\ast}$    &  0.43 \\
WD 1046$-$017               & $-$0.13(0.02)          & -                      & -                      &  - \\
WD 1448$+$411               & $-$0.12(0.01)          & -                      & 0.30(0.13)             &  - \\
SDSS J1238$+$2910           & $-$0.14(0.04)          & -                      & 0.85(0.17)             &  - \\
WD 2207$+$121               & $-$0.19(0.04)          & -                      & 0.65(0.18)             &  0.74 \\
SDSS J0738$+$1835           & $-$0.12(0.04)          & -                      & -                      &  - \\
Gaia 2204$+$0233            & $-$0.15(0.02)          & -                      & 0.15(0.41)$^{\ast}$    &  - \\
WD 1145$+$017               & $-$0.19(0.02)          & 1.06(0.33)             & 0.49(0.39)             &  - \\
WD 0902$+$399               & $-$0.09(0.03)          & -                      & -                      &  - \\
Gaia 0155$+$0431            & $-$0.12(0.02)          & -                      & $-$0.14(0.11)          &  - \\
WD 1551$+$175               & $-$0.18(0.02)          & -                      & -                      &  0.19 \\
GD 56                       & $-$0.16(0.01)          & 2.98(0.16)             & 0.89(0.06)             &  2.94 \\
SDSS J0234$-$0406           & $-$0.09(0.01)          & -                      & 0.26(0.18)             &  - \\
HE 0106$-$3253              & $-$0.20(0.01)          & 0.78(0.24)             & 0.14(0.11)             &  0.08 \\
WD 0842$+$572               & $-$0.18(0.01)          & 3.30(0.18)             & 0.72(0.05)             &  - \\
GD 61                       & $-$0.23(0.01)          & 0.60(0.16)             & 0.21(0.09)             &  0.28 \\
HE 1349$-$2305              & $-$0.21(0.02)          & -                      & 0.48(0.14)             &  0.33 \\
SDSS J0347$+$1624           & $-$0.19(0.01)          & -                      & 0.42(0.14)             &  0.38 \\
SDSS J1043$+$0855           & $-$0.17(0.04)          & 0.73(0.51)             & $-$0.29(0.53)$^{\ast}$ &  0.15 \\
Gaia J0611$-$6931           & $-$0.19(0.02)          & -                      & 0.79(0.05)             &  5.16 \\
Gaia J0644$-$0352           & $-$0.29(0.01)          & -                      & 0.57(0.14)             &  0.49 \\
WD 0420$-$731               & $-$0.24(0.01)          & 1.89(0.21)             & 0.60(0.06)             &  - \\
WD J0529$-$3401             & $-$0.27(0.02)          & -                      & 0.83(0.07)             &  - \\
WD 1015$+$161               & $-$0.27(0.01)          & 0.90(0.10)             & 0.50(0.12)             &  0.17 \\
SDSS J1058$+$0206           & $-$0.21(0.03)          & 0.40(0.47)$^{\ast}$    & 0.52(0.59)$^{\ast}$    &  - \\
WD J2212$-$1352             & $-$0.26(0.04)          & -                      & 0.74(0.15)             &  - \\
WD 1457$-$086               & $-$0.20(0.03)          & 0.04(0.47)$^{\ast}$    & $-$0.19(0.30)$^{\ast}$ &  0.04 \\
WD 0110$-$565               & $-$0.29(0.01)          & 0.73(0.31)             & 0.43(0.12)             &  0.15 \\
SDSS J0845$+$2257           & $-$0.29(0.01)          & 1.24(0.15)             & 0.44(0.11)             &  0.6 \\
PG 1031$+$063               & $-$0.27(0.02)          & 0.41(0.29)             & 0.32(0.36)$^{\ast}$    &  - \\
SDSS J1228$+$1040           & $-$0.29(0.02)          & 1.97(0.10)             & 0.70(0.12)             &  0.4 \\
WD 2328$+$107               & $-$0.24(0.01)          & 0.12(0.18)$^{\ast}$    & 0.22(0.24)$^{\ast}$    &  0.05 \\
Gaia J0510$+$2315           &   $-$0.33(0.01)        & -                      & 0.50(0.10)             &  - \\
WD 1622$+$587               &   $-$0.29(0.02)        & -                      & 0.50(0.09)             &  0.27 \\
WD 1018$+$410               &   $-$0.32(0.01)        & -                      & 0.31(0.24)             &  0.11 \\
WD 0420$+$520               &   $-$0.33(0.01)        & 1.54(0.22)             & 0.42(0.08)             &  - \\
WD 0843$+$516               &   $-$0.32(0.01)        & 1.08(0.33)             & 0.45(0.14)             &  0.13 \\
SDSS J0006$+$2858           &   $-$0.34(0.02)        & -                      & 0.61(0.08)             &  - \\
EC 05365$-$4749             &   $-$0.35(0.01)        & 1.04(0.23)             & 0.51(0.08)             &  - \\
WD 1929$+$011               &   $-$0.33(0.01)        & 1.66(0.08)             & 0.45(0.06)             &  0.19 \\
WD J2133$+$2428             &   $-$0.32(0.02)        & -                      & 0.35(0.09)             &  - \\
Gaia J2100$+$2122           &   $-$0.40(0.02)        & 2.27(0.16)             & 0.71(0.06)             &  0.39 \\
PG 0010$+$281               &   $-$0.38(0.01)        & 0.47(0.32)             & 0.22(0.16)             &  - \\
SDSS J0928$+$1332           &   $-$0.33(0.03)        & -                      & 0.85(0.38)             &  - \\
WD 1454$+$172               &   $-$0.45(0.02)        & -                      & 0.74(0.28)             &  - \\
\end{longtable}
\vspace{-5mm}
\noindent Fractional IR Luminosities, ${\rm {L_{IR}/L_*}}$, were obtained from \citet{2015MNRAS.449..574R} and \citet{2020ApJ...905....5D}.\\
\noindent $^{\ast}$ Stars discarded due to large errors in colors. \\
\noindent $^{\ast \ast}$ Objects excluded because of possible contamination of nearby sources detected by visual inspection of optical and infrared images.
\end{appendices}

\bibliographystyle{rmaa}
\bibliography{reference-file}

\end{document}